\documentclass{amsart}
\usepackage{graphicx}
\theoremstyle{definition}

\theoremstyle{remark}

\numberwithin{equation}{section}

\begin{document}

\title[REALISTIC INTERATOMIC POTENTIAL FOR MD SIMULATIONS]
{REALISTIC INTERATOMIC POTENTIAL FOR MD SIMULATIONS}%
\author{Yu.V.Eremeichenkova, L.S.Metlov, A.F.Morozov}%
\address{Donetsk Institute of Physics and Engineering, 72, R. Luxembourg str.,
83114 Donetsk, Ukraine}%
\email{erem$_{-}$yulia@atlasua.net, metlov@atlasua.net}%


\begin{abstract}
The coefficients of interatomic potential of simple form Exp-6 for
neon are obtained. Repulsive part is calculated ab-initio in the
Hartree-Fock approximation using the basis of atomic orbitals
orthogonalized exactly on different lattice sites. Attractive part
is determined empirically using single fitting parameter. The
potential obtained describes well the equation of state and
elastic moduli of neon crystal in wide range of interatomic
distances and it is appropriate for molecular dynamic simulations
of high temperature properties and phenomena in crystals and
liquids.
\end{abstract}
\maketitle
\section{Introduction}

Investigation of strongly anharmonic nonlinear atomic systems by
molecular dynamics (MD) method at high temperatures, pressures, or
study of systems effected by large amplitude excitations requires
high accuracy of interatomic potential (IP). Series expansion of
the IP in the displacements of atoms from equilibrium positions is
widely used both in phonon theory and in MD simulation
\cite{m1,m2}. Usually, fourth-order anharmonisms or lower-order
ones can be taken into account because of complexity of expansion
coefficients calculation. As an alternative, realistic potential
method is used \cite{m3,m4,m5}, in which exact equations of motion
of atoms are solved using IP of concrete substance without series
expansion. Owing to that, all-order anharmonisms are taken into
account automatically. This advantage of realistic potential
method is especially useful in the MD simulation of soliton
solutions where atoms approach each other closely. Realistic IP
should have simplest form to reduce calculation expenses as well
as it must describe precisely the properties of the substance
under extreme conditions. The aim of this paper is to obtain such
IP.

Conventional way of realistic IP determination is empirical
fitting to the properties of gas or a crystal near the equilibrium
point \cite{m6,m7}. However, such potentials become unreliable at
small interatomic distances like to that arising in soliton waves.
The properties of highly compressed matter (e.g., for neon up to 1
Mbar \cite{m8}) could give an information for obtaining
all-distance reliable IP. However, the set of properties, which
can be measured accurately at megabar pressures is restricted
strongly. Practically, the equation of state and bulk modulus only
may be included in this set \cite{m8,m9}. For $C_{ik}$ modules the
precision worsens drastically even at kilobar pressures
\cite{m10,m11}. There is insufficiency of empirical information
for fitting all the parameters of IP, and ab-initio calculation is
required.

Realistic IP via interatomic distance is obtained in present work
for the crystal and dimer of neon. Repulsive part of the potential
is calculated ab-initio in Hartree-Fock approximation using the
basis of localized atomic orbitals orthogonalized exactly on
different lattice sites. Attractive part is chosen to have
standard Van-der-Vaals form of $Cr^{-6}$ with single empirical
parameter $C$. Used approximations and calculation details are
described in the section 2. In the section 3 repulsive part of IP
is interpolated by exponential function of interatomic distance
(Exp-6 potential) and the IP parameter are determined.
Experimental verification of the IP obtained is performed in the
section 4 using the data concerning equation of state \cite{m8,m9}
and elastic moduli \cite{m12,m13,m14,m15,m16,m17} of compressed
neon. The IP calculated is found to be in a good consistence with
the experiment in whole range of pressure.

\section{Ab-initio calculation of repulsion potential}

In MD simulations by realistic potential method the problem is
divided into two stages. The former is quantum-mechanical
calculation of the IP at electron level, with interatomic distance
considering as a parameter. The latter is solving equations of
motion of atoms using the IP obtained. This division is
correspondent to adiabatic approximation when atoms and electrons
motion is described separately \cite{m18}.

Since pair collisions of atoms have maximal probability, we
concentrate the attention on the dimer of neon, and define the IP
as a cohesive energy of the dimer. Three-atom forces can be taken
into account as a correction to the two-atom ones using
incremental expansion \cite{m19}. The estimation of \cite{m19}
shows three-atom force contribution to be small.

In Hartree-Fock approximation short-range repulsive part of IP is
expressed through one-electron density matrix. We don't use hard
core approximation. Rearrangement of all electron shells is
allowed as interatomic distance is altered.

Localized basis of atomic orbitalls orthogonalized exactly (by
Lovdin procedure \cite{m20}) on different lattice sites is used.
In this basis one-electron density matrix has the form \cite{m21}

\begin{eqnarray}\label{e:1}
\nonumber   \rho(\textbf{r}'|\textbf{r}; \{\textbf{l}\})= 2
\sum_{\textbf{l}s} \{\varphi_{s}(\textbf{r}'-\textbf{l})
\varphi^{*}_{s}(\textbf{r}-\textbf{l}) - \sum_{\textbf{l}'s'}
\varphi_{s'}(\textbf{r}'-\textbf{l}')
P^{\textbf{l}'\textbf{l}}_{s's}\varphi^{*}_{s}(\textbf{r}-\textbf{l})  \},  \\
\textbf{P}=\textbf{I}-(\textbf{I}+\textbf{S})^{-1},
\end{eqnarray}
where $\varphi_{s}(\textbf{r}-\textbf{l})=|\textbf{l}s>$ is wave
function of electron in isolated atom (atomic orbital),
$\textbf{l}$ and $\textbf{l}'$ are radius-vectors of lattice
sites, $s$ numerates occupied states of the atom, $\textbf{P}$ is
orthogonalizing matrix, $\textbf{I}$ is unit matrix, $\textbf{S}$
is overlap integral matrix with the elements

\begin{eqnarray}\label{e:2}
\nonumber
S^{\textbf{l}'\textbf{l}}_{s's}=<\textbf{l}'s'|\textbf{l}s>;
\textbf{l}\neq\textbf{l}',  \\
 S^{\textbf{l}'\textbf{l}}_{s's}=0; \textbf{l}=\textbf{l}'.
\end{eqnarray}

We expand repulsive part of IP in the terms of small parameter
such as the largest overlap integral $S$. Usually, $S<<1$ in
uncompressed crystal, and overlap integrals grow exponentially as
interatomic distance is decreased. The IP is expressed through the
products of elements of orthogonalizing matrix $\textbf{P}$ and
two-center Slater-Koster integrals. These integrals are atomic
obital matrix elements of crystal hamiltoinial operators. The
order in $S$ for two-center integrals is estimated using the
theorem about average value. The elements of matrix
$\textbf{P}=\textbf{I}-(\textbf{I}+\textbf{S})^{-1}$ are expanded
in powers of overlap integrals matrix $\textbf{S}$

\begin{eqnarray}\label{e:3}
\nonumber
P^{\textbf{l}\textbf{l}'}_{ss'}=S^{\textbf{l}\textbf{l}'}_{ss'}+O(\textbf{S}^{2}),\\
P^{\textbf{l}\textbf{l}}_{ss'}=-
(\textbf{S}^{2})^{\textbf{l}\textbf{l}}_{ss'}+O(\textbf{S}^{3}).
\end{eqnarray}

The elements of matrix $\textbf{P}$ contain high-order terms along
with the main ones proportional to $S$ and $S^2$.

Using the estimations described above, we expand the repulsive
part of IP in powers of $S$

\begin{eqnarray}\label{e:4}
V_{sr}=E^{(0)}+W_2+W_4+W_6.
\end{eqnarray}
Here $E^{(0)}$ is the energy of interatomis interaction if
orthogonalizing of neighbor atoms orbitals is neglected, $W_2,
W_4, W_6$ are orthogonalizing corrections. Series expansion in $S$
begins for them from the second, the third, and the sixth powers
respectively. Due to the presence of matrix $\textbf{P}$,
orthogonalizing corrections contain high-order terms in $S$ along
with the main ones.

In the equation \ref{e:4}

\begin{eqnarray}\label{e:5}
E^{(0)}=\sum_{\textbf{l}s}
\sum_{\textbf{m},\textbf{m}\neq\textbf{l}}
\langle\textbf{l}s|V^{\textbf{m}}_{en}+V^{\textbf{m}}_{a}+V^{\textbf{m}}_{ex}|
\textbf{l}s\rangle +U_{nn}.
\end{eqnarray}
The first term in equation \ref{e:5} consists of two-center
integrals. They are atomic orbital matrix elements of electron-ion
interaction potential $V^{\textbf{m}}_{en}$, of neutral isolated
atom potential $V^{\textbf{m}}_{a}$, of electron-electron exchange
interaction potential $V^{\textbf{m}}_{ex}$ respectively. The
second term is the energy of nucleus-nucleus interaction.
Electron-ion interaction potential has the form

\begin{eqnarray}\label{e:6}
V^{\textbf{m}}_{en}=V_{en}(\textbf{r}-\textbf{m})=-Ze^{2}/
|\textbf{r}-\textbf{m}|.
\end{eqnarray}
Neutral isolated atom potential is

\begin{eqnarray}\label{e:7}
V^{\textbf{m}}_{a}=V_{a}(\textbf{r}-\textbf{m})=V_{en}(\textbf{r}-\textbf{m})
+ 2 \sum_{t}   <\textbf{m}t|v_c|\textbf{m}t>,
\end{eqnarray}
where

\begin{eqnarray}
\nonumber
<\textbf{m}t|v_{c}|\textbf{m}t>=\int\varphi^{*}_{t}(\textbf{r}'
-\textbf{m})
v_{c}(\textbf{r}-\textbf{r}')\varphi_{t}(\textbf{r}'-\textbf{m})d\textbf{r}' ,     \\
\nonumber
v_{c}(\textbf{r}-\textbf{r}')=e^{2}/|\textbf{r}-\textbf{r}'|.
\end{eqnarray}
Action of electron-electron exchange interaction potential on wave
function is defined as

\begin{eqnarray}\label{e:8}
<\textbf{l}s|V_{ex}^{\textbf{m}}|\textbf{l}s>=
-\sum_{t}<\textbf{l}s,\textbf{m}t|v_{c}|\textbf{l}s,\textbf{m}t>.
\end{eqnarray}

In the equation \ref{e:4} orthogonalizing corrections, $W_{2},
W_{4}, W_{6}$, are of the form

\begin{eqnarray}\label{e:9}
\nonumber  W_2=-2  \sum_{\textbf{l}s}
\sum_{\textbf{l}'s',\textbf{l}\neq\textbf{l}'}
P_{ss'}^{\textbf{l}\textbf{l}'}
<\textbf{l}'s'|V^{\textbf{l}'}_{a}+V^{\textbf{l}'}_{ex}|\textbf{l}s> -\\
- \sum_{\textbf{lm}s s'tt',\textbf{l}\neq\textbf{m}}
P_{ss'}^{\textbf{ml}}   P_{tt'}^{\textbf{lm}}
<\textbf{l}s',\textbf{m}t'|v_{c}|\textbf{m}s,\textbf{l}t>;
\end{eqnarray}

\begin{eqnarray}\label{e:10}
\nonumber W_{4}=\sum_{\textbf{l}s s'tt'}
P_{ss'}^{\textbf{l}\textbf{l}}  P_{tt'}^{\textbf{l}\textbf{l}}
\{2<\textbf{l}s',\textbf{l}t'|v_{c}|\textbf{l}t,\textbf{l}s> -
<\textbf{l}s',\textbf{l}t'|v_{c}|\textbf{l}s,\textbf{l}t>\} -  \\
\nonumber
-2\sum_{\textbf{l}s s'}  P_{ss'}^{\textbf{l}\textbf{l}}
<\textbf{l}s'|\sum_{\textbf{m}\neq\textbf{l}}
(V^{\textbf{m}}_{a}+V^{\textbf{m}}_{ex})|\textbf{l}s> +\\
\nonumber
  + 2 \sum_{\textbf{l}\textbf{m}s
s'tt',\textbf{l}\neq\textbf{m}}
 \{P_{ss'}^{\textbf{l}\textbf{l}} P_{tt'}^{\textbf{m}\textbf{m}}
<\textbf{l}s',\textbf{m}t'|v_{c}|\textbf{m}t,\textbf{l}s> +
P_{ss'}^{\textbf{m}\textbf{l}} P_{tt'}^{\textbf{l}\textbf{m}}
<\textbf{l}s',\textbf{m}t'|v_{c}|\textbf{l}t,\textbf{m}s>\} +  \\
\nonumber + 2 \sum_{\textbf{l}\textbf{m}s
s'tt',\textbf{l}\neq\textbf{m}} P_{ss'}^{\textbf{m}\textbf{l}}
P_{tt'}^{\textbf{m}\textbf{l}}
\{2<\textbf{l}s',\textbf{l}t'|v_{c}|\textbf{m}t,\textbf{m}s> -
<\textbf{l}s',\textbf{l}t'|v_{c}|\textbf{m}s,\textbf{m}t>\} +\\
+ 4\sum_{\textbf{l}\textbf{m}s s'tt',\textbf{l}\neq\textbf{m}}
P_{ss'}^{\textbf{m}\textbf{l}} P_{tt'}^{\textbf{l}\textbf{l}}
\{2<\textbf{l}s',\textbf{l}t'|v_{c}|\textbf{l}t,\textbf{m}s> -
<\textbf{l}s',\textbf{l}t'|v_{c}|\textbf{m}s,\textbf{l}t>\};
\end{eqnarray}

\begin{eqnarray}\label{e:11}
 W_{6}=-\sum_{\textbf{l}\textbf{m}s s'tt',\textbf{l}\neq\textbf{m}}
P_{ss'}^{\textbf{l}\textbf{l}}P_{tt'}^{\textbf{m}\textbf{m}}
<\textbf{l}s',\textbf{m}t'|v_{c}|\textbf{l}s,\textbf{m}t>.
\end{eqnarray}

Since the orthogonalizing corrections grow exponentially as the
interatomic distance is decreased it is impossible to say what
correction may be neglected. It should be checked for each
substance under consideration.

Using the method described, we calculate repulsive part of IP,
$V_{sr}$ (equation \ref{e:4}), for neon dimer as a function of
interatomic distance $d$. Atomic orbitals from Clementi-Roetti set
\cite{m22} are used as a basis. Hartree system of atomic units
$\hbar=e=m_e=1$ is applied. The calculation shows the terms
$E^{(0)}$ and $W_2$ in equation \ref{e:4} to have the same order
of magnitude and opposite signs. These terms are found to give
major contributions to the IP. The $W_4$ correction consists of
0,02 per cent of the IP at equilibrium interatomic distance $d_0$.
Further, the $W_4$ does not exceed of 1 per cent of the IP up to
$d\sim0.75 d_0$. Finally, at small $d$, like to that arising in
soliton waves (for $d$ above 0,6-0.75 $d_0$), the $W_4$ becomes
about 2-4 per cent of the IP. The contribution of $W_6$ to the IP
is small negligibly (0.002 per cent) in whole range of $d$ under
consideration.

\section{Determination of interatomic potential parameters}

We interpolate calculated points $V_{sr}(d)$ by exponential
function of interatonic distance using least square method by the
formula

\begin{eqnarray}\label{e:12}
\nonumber V_{sr}(d) = A_0 \exp(-\alpha(x-1));\\ x=d/z_0
\end{eqnarray}
with two unknown parameters $A_0$ and $\alpha$. Experimental
equilibrium interatomic distance for neon dimer $z_0=5,8411$ a.u.
\cite{m23} is used as the third parameter of the IP. The
parameters are found to be $A_0=(1,1384\pm0,0002)\cdot10^{-4}$
a.u., $\alpha=13,6407\pm0,0037$. Interpolation error is 4-1 per
cent of $V_{sr}$ when the $d$ is altered from equilibrium one to
0.6$z_0$.

Adding the attractive part, we express the IP in standard Exp-6
form

\begin{eqnarray}\label{e:13}
\nonumber V(d) = A_0 \exp(-\alpha(x-1)) - Cd^{-6};\\x=d/z_0.
\end{eqnarray}

A single unknown parameter $C$ remains in attractive part of IP.
We propose to fit the $C$ to experimental equilibrium interatomic
distance. Using of equilibrium data is considered to be reliable
at all interatomic distances since the attraction is essential
near the equilibrium only while ab-initio calculated repulsive
part dominates at small $d$.

For MD simulation of lattice dynamics, it is possible to fit the
$C$ to experimental data for dimer at $T$=0 K because the
temperature effects will be taken into account explicitly, at the
stage of equations of motion solving. In this case, for neon
$C$=10,7293 (experimental equilibrium interatomis distance in the
dimer is $z_0$=5,8411 a.u. \cite{m23}). Calculated cohesive energy
of dimer is $E_{coh}=-1,4497\cdot10^{-4}$ a.u., experimental one
is $E_{coh}=-1,338\cdot10^{-4}$ a.u. \cite{m23}. The discrepancy
is 7 per cent of experimental value.

For calculating static properties of a crystal at finite
temperature, e.g., equation of state, elastic modules, it is
better to fit the $C$ to experimental data for a crystal at the
same temperature. Such determination allows one to take into
account implicitly three-atom forces, temperature effects,
zero-point oscillations, and other effects omitted at the stage of
IP calculating. In this case, for neon $C$=7,4030 (experimental
equilibrium interatomis distance in the crystal is $d_0$=5,9647
a.u. at $T=4,25$ K \cite{m12}). Calculated cohesive energy of
uncompressed crystal is $E_{coh} =-6,7620\cdot10^{-4}$ a.u. per
atom, experimental one is $E_{coh}=-(7,35\pm0,03)\cdot10^{-4}$
a.u. \cite{m24}. The discrepancy is 7.6 per cent of experimental
value.

\section{Results and discussion}

\begin{figure}[p]
  \includegraphics[width=5.5in]{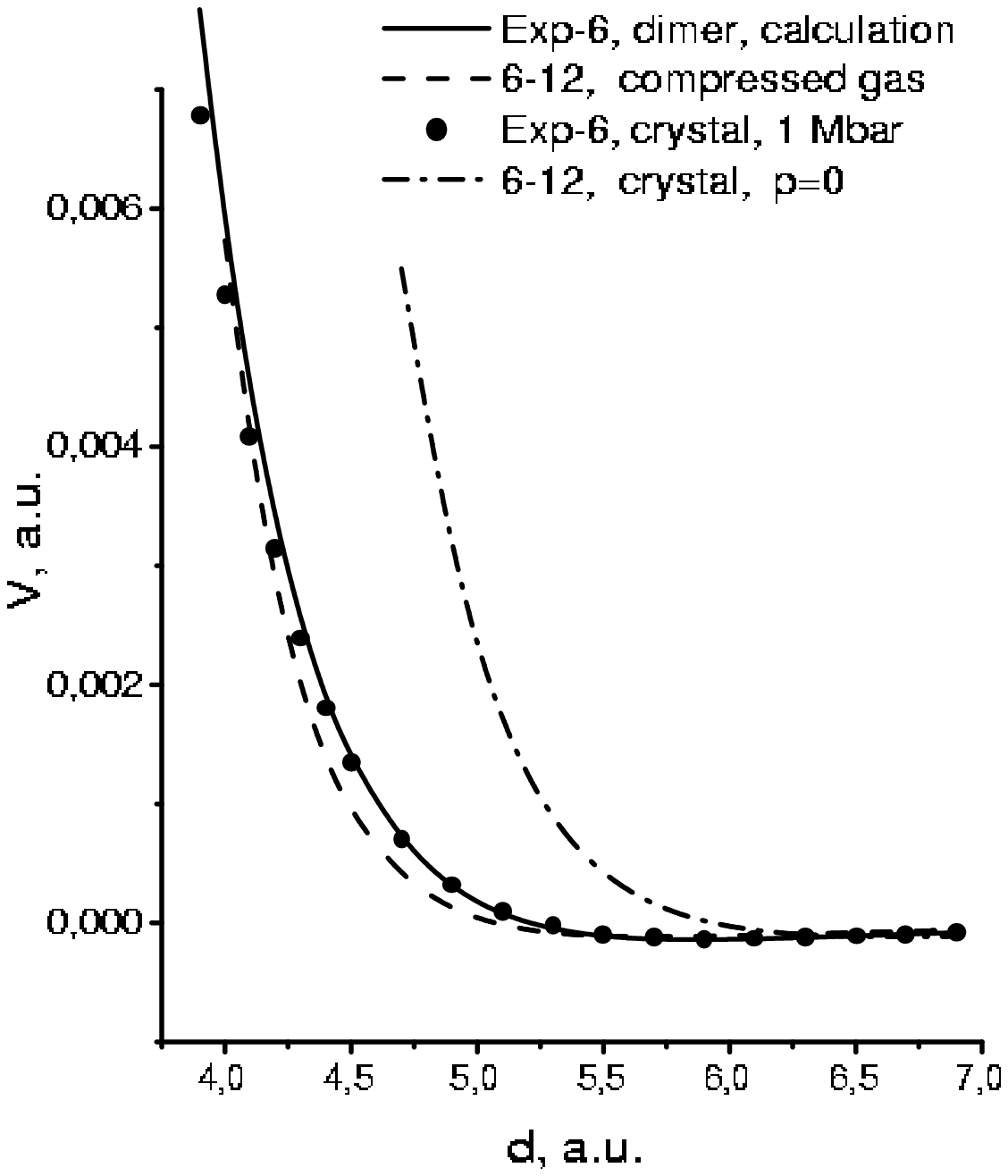}\\
  \caption{Calculated IP and three potentials fitted to experimental
  data for neon (from \protect\cite{m6,m8,m7} respectively).}\label{f1}
\end{figure}

Interatomic potential of neon is given in the figure \ref{f1} as a
function of interatomic distance $d$. The IP calculated by
equation \ref{e:13} for dimer is plotted by solid curve.
Van-der-Vaals constant ($C$=10,7293) is fitted to experimental
equilibrium interatomic distance in dimer \cite{m23}.

"Experimental" IP obtained in \cite{m8} is denoted by solid
circles. This IP had been determined by interpolating experimental
data $p(V)$ (measured at 300 K) by the formula Exp-6. The
interpolation had been performed in theoretical model taking
thermal pressure and zero-point oscillations into account
explicitly, excluding them from the definition of IP. It allows us
to compare the 300 K data of \cite{m8} with our zero-temperature
result. Three-atom forces didn't include explicitly in the model
of \cite{m8}. However, in \cite{m8}, the effect of these forces is
taken into account implicitly through fitting the IP to
experimental data for a crystal. In our calculation three-atom
forces are omitted because of fitting to dimer data. The agreement
of calculated IP and experimental one indicates that three-atom
forces in neon are small at the pressures up to 1Mbar.

Two remaining curves in the figure \ref{f1} are interatomic
potentials of neon obtained by fitting to experimental data using
Lennard-Jones potential (6-12 formula)

\begin{eqnarray}
\nonumber V(x)=\varepsilon(-2/x^6 + 1/x^{12})\\ \nonumber x=d/z_0,
\end{eqnarray}
where $\varepsilon$ and $z_0$ are fitting parameters. Dashed curve
is the IP obtained using corresponding-states law fitted to
vapor-pressure ratio of isotopic liquid \cite{m6}. Dashed-dotted
curve is the IP fitted to experimental lattice constant and
cohesive energy of crystal neon at $p=0$, $T=0$ K \cite{m7}.
Fitting to equilibrium crystal properties leads to bad describing
the IP for compressed crystal. Fitting to compressed gas
properties gives the values of the IP close to experimental ones
at moderated pressures.

\begin{figure}[p]
  \includegraphics[width=5.5in]{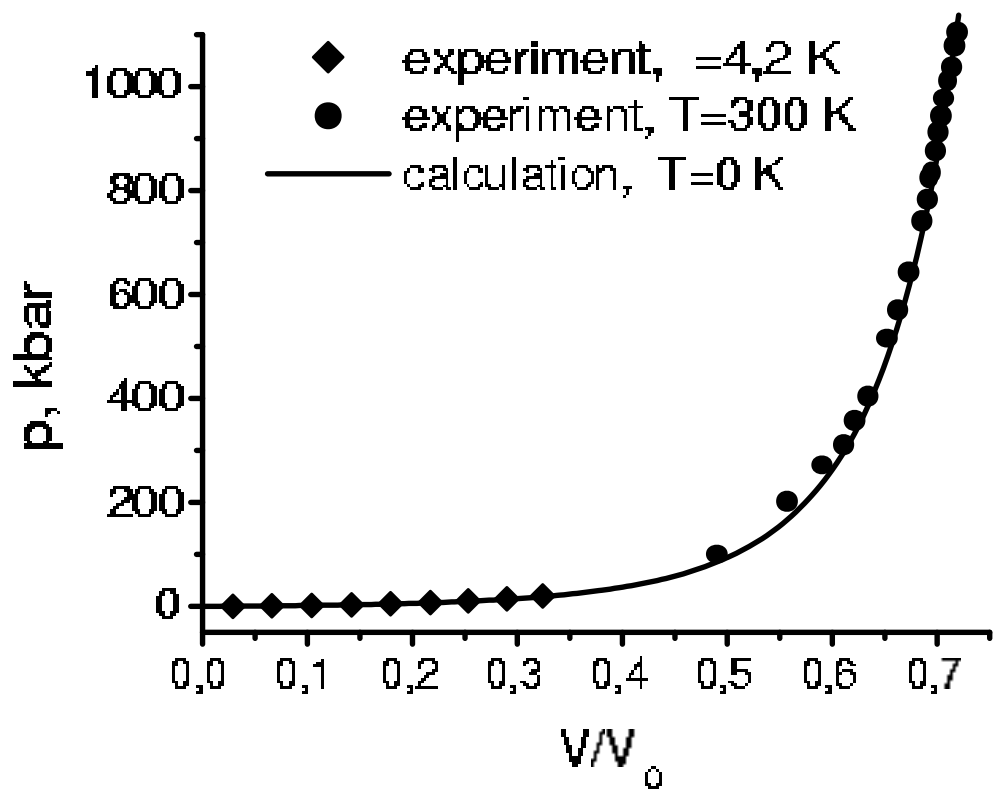}\\
  \caption{Equation of state for solid neon (experiment from \protect\cite{m9} (4.2 K)
  and \cite{m8} (300 K)).}\label{f2}
\end{figure}

Using the IP obtained (eq. \ref{e:13}) we calculate the equation
of state $p(V)$ for solid neon. Calculated pressure p against
fractional volume is given in the figure \ref{f2} as solid curve.
Van-der-Vaals constant is fitted to experimental equilibrium
interatomic distance $d_0$=5,9647 a.u. measured for crystal neon
at $T$=4.25 K, $p$=0 \cite{m12}.

Experimental points $p(V)$ from \cite{m8} ($T$=300 K) and
\cite{m9}  ($T$=4.2 K) are also given in the figure \ref{f2}. At
the pressures below 20 kbar theoretical curve is in a good
agreement with the experimental points of \cite{m9}. At moderated
pressures theoretical curve deviates from experimental points of
\cite{m8} by 4 per cent. This deviation caused, mainly, by
neglecting of thermal pressure in our calculation. Figure \ref{f2}
shows temperature sensitivity of the equation of state to be
small.

\begin{figure}[p]
  \includegraphics[width=5in]{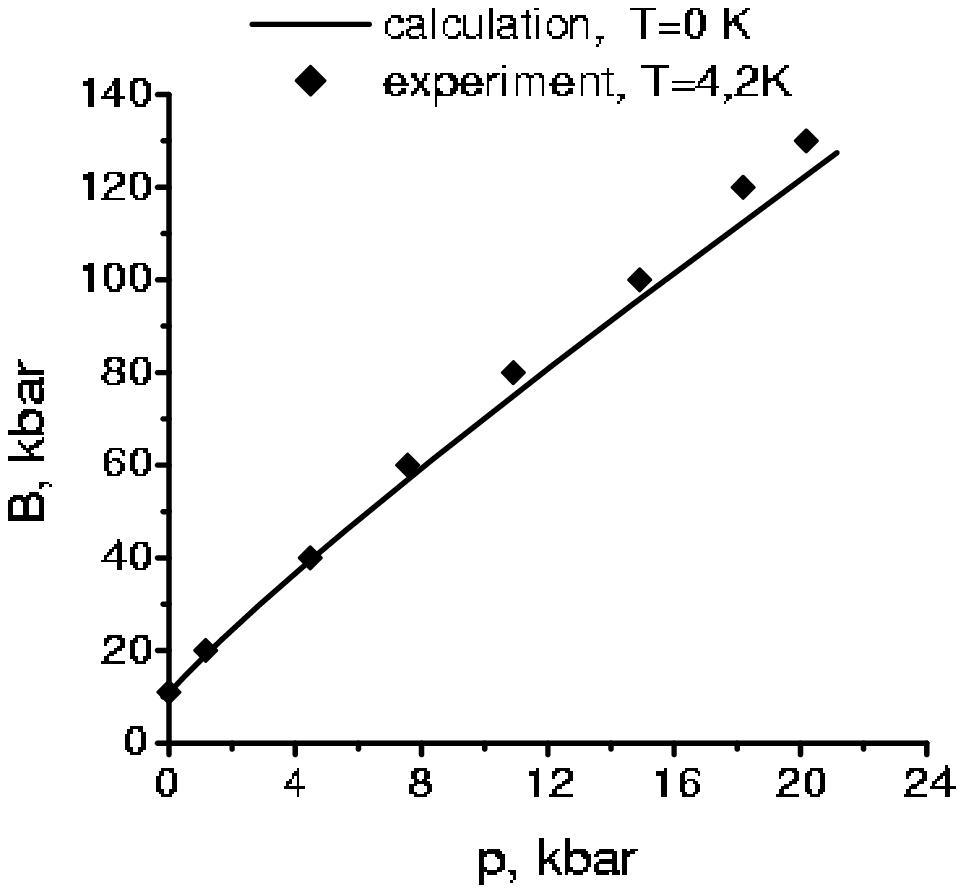}\\
  \caption{Bulk modulus of solid neon (experiment from \protect\cite{m9}).}\label{f3}
\end{figure}

We calculate bulk modulus of solid neon by means of the IP
obtained. Van-der-Vaals constant is fitted to experimental
equilibrium interatomic distance in the crystal \cite{m12}.
Calculated bulk modulus $B$ via the pressure $p$ is given in the
figure \ref{f3} as solid curve. Experimental points obtained in
\cite{m9} at $T$=4.2 K are plotted as solid symbols. Bulk modulus
is seen to be more sensitive to the approximations used. Growing
when the $p$ is enhanced, the difference between calculated $B$
and measured one becomes about 7 per cent  of experimental $B$ at
$p$=20 kbar. Incorrect taking three-atom forces into account at
moderated pressures is seems to contribute mainly in this
discrepancy. In our calculation three-atom forces (and zero-point
oscillations too) are taken into account implicitly, by fitting
the IP to experimental data for uncompressed crystal. Thus,
calculated $B$ agrees with experimental one at small pressures
only (to 8 kbar). One can't determine correctly the dynamics of
alteration of three-atom forces with enhancing of pressure. It is
the cause of growing the deviation of calculated $B$ from measured
one.

We calculate elastic modules $C_{ik}$ using the IP obtained with
Van-der-Waals constant fitted to crystal experimental data
\cite{m12}. Calculated modules and experimental ones are given in
table \ref{T1} for uncompressed solid neon at low temperatures.
Isothermic modules had been obtained in static measurements
\cite{m9,m12}. Adiabatic modules had been measured in ultrasonic
and neutron scattering experiments \cite{m13,m14,m15,m16,m17}.
However, the difference between isotermic modules and adiabatic
ones is negligible at the temperatures under consideration (see,
e.g., \cite{m13}).

\begin{table}
\caption{\label{T1}Elastic modules of solid neon}
\begin{tabular}{|c|c|c|c|c|c|c|c|}
  \hline
    Ref. & $T$, K & $B$, kbar & $C_{11}$, kbar & $C_{12}$, kbar & $C_{44}$, kbar &
  $\delta= \frac{(C_{44}-C_{12})}{C_{12}}$ & Method \\
  \hline
  \cite{m12} & 4,25& 11,12$\pm$0,12 & - & - & - & - & Static mea- \\
  \cite{m9}  & 4,2 & 11,0$\pm$0,1   & - & - & - & - & surements $p(V)$\\
  \hline
  \cite{m13} & 4   & 11,36$\pm$0,26 & - & - & - & - & Ultrasonic ve-\\
  \cite{m14} & 5   & 11,2$\pm$0,5   & - & - & - & - & locity measur.\\
  \hline
  \cite{m15} & 4,7 & 12,1$\pm$0,4  & 16,9$\pm$0,5  & 9,7$\pm$0,4 & 10,0$\pm$0,3& 0.03$\pm$0,07 &
                                                               Inelastic neu-  \\
  \cite{m16} & 5   & 11,24$\pm$0,17& 16,61$\pm$0,17&8,55$\pm$0,21& 9,52$\pm$0,05& 0.11$\pm$0,03&
                                                                             tron-phonon\\
  \cite{m17} & 6   & 11,52$\pm$0,3 & 16,49$\pm$0,3 & 9,03$\pm$0,3& 9,28$\pm$0,08& 0.03$\pm$0,04&
                                                                              scattering\\
  \hline
  Calc.      & 0   & 10,76       & 14,95       & 8,67 & 8,67 & 0 & Ab-initio calc.\\
  \hline
\end{tabular}
\end{table}

The $C_{ik}$ modules are seen to be more sensitive to the
measurement method and calculation approximations. The difference
between theoretical and experimental values of $C_{11}$ and
$C_{44}$ is about 10 per cent of experimental values for most
accurate experiment \cite{m16}. The agreement is better for
$C_{12}$ modulus (the discrepancy is about 2 per cent \cite{m16}).
The deviation from Cauchy relation $\delta=(C_{44}-C_{12})/C_{12}$
is also given in table \ref{T1}. Cauchy violation is the measure
of deviation of the IP from spherical symmetry. The
$\delta=0,11\pm0,03$ in \cite{m19}, while it falls into
experimental error bar in other experiments listed in the table
\ref{T1}. Cauchy relation takes place for our calculation results
because spherical symmetry form of the IP is supposed in
theoretical model. Small value of experimental $d$ indicates that
spherical symmetry approximation for IP is valid for uncompressed
neon at least. For another rare gas crystal, krypton, experiment
\cite{m11} shows Cauchy relation to satisfy well under pressure up
to 80 kbar. Moreover, for MgO the Cauchy violation is measured to
drop with enhancing pressure up to 200 kbar \cite{m25}.

Unlike to $C_{ik}$ modules, bulk modulus $B$ is less sensitive to
measurement method and calculation approximations. The discrepancy
of theoretical result and experimental one doesn't exceed of 4 per
cent and falls into experimental error frames.

\section{Conclusion}

Coefficients of realistic IP of simple form Exp-6 are obtained for
neon by ab-initio calculation of repulsive part in Hartree-Fock
approximation in the basis of atomic orbitals orthogonalized
exactly on different lattice sites. Attractive part is determined
empirically using single fitting parameter, Van-der-Vaals constant
$C$. For fitting the $C$ it is enough to know experimental
equilibrium interatomic distance in crystal (or dimer), i.e. high
pressure experimental data is not required. The IP calculated is
suitable for molecular dynamic simulations of high temperature and
high pressure properties and phenomena in crystals and liquids due
to simplicity of the form and precise describing experimental data
in wide range of interatomic distances.

\bibliographystyle{amsplain}
\bibliography{Erem_IP}

\providecommand{\bysame}{\leavevmode\hbox to3em{\hrulefill}\thinspace}
\providecommand{\MR}{\relax\ifhmode\unskip\space\fi MR }
\providecommand{\MRhref}[2]{%
  \href{http://www.ams.org/mathscinet-getitem?mr=#1}{#2}
}
\providecommand{\href}[2]{#2}
\begin{thebibliography}{10}

\bibitem{m21}
I.V. Abarenkov and I.M. Antonova, Phys. Stat. Sol. \textbf{38} (1970), 783.

\bibitem{m9}
M.S. Anderson and S.A. Swenson, J. Phys. Chem. Sol. \textbf{36} (1975), 145.

\bibitem{m14}
R.~Balzer, D.S. Kupperman, and R.O. Simmons, Phys. Rev. B. \textbf{10} (1971),
  no.~3636.

\bibitem{m12}
D.N. Batchelder, D.L. Losee, and R.O. Simons, Phys. Rev. \textbf{162} (1967),
  767.

\bibitem{m13}
P.A. Bezugly, R.O. Plakhotin, and L.M. Tarasenko, Fiz. Tv. Tela (Sov.)
  \textbf{12} (1970), 1199.

\bibitem{m2}
S.~R. Bichkham, S.A. Kiselev, and A.J. Sievers, Phys. Rev. B \textbf{47}
  (1993), 14206.

\bibitem{m6}
G.~Boato and G.~Casanova, Physica \textbf{27} (1961), 571.

\bibitem{m18}
M.~Born and K.~Huang, \emph{Dynamical theory of crystal lattices}, Oxford:
  Clarendon, 1954.

\bibitem{m7}
J.S. Brown, Proc. Phys. Soc. (London) \textbf{89} (1966), 987.

\bibitem{m22}
E.~Clementi and C.~Roetti, \emph{Atom data nucl. data table}, vol.~14, p.~177,
  1974.

\bibitem{m1}
T.~Cretegny, T.~Dauxois, and S.~Ruffo, Physica D \textbf{121} (1998), 109.

\bibitem{m17}
Y.~Endoh, G.~Shirane, and J.~Jr. Skalyo, Phys. Rev. B \textbf{11} (1975), 1681.

\bibitem{m8}
R.J. Hemley, C.S. Zha, H.K. Mao, A.P. Jephcoat, L.W. Finger, and D.F. Cox,
  Phys. Rev. B \textbf{39} (1989), 11820.

\bibitem{m15}
J.A. Leake, W.B. Daniels, J.~Jr. Skalyo, B.C. Frazer, and G.~Shirane, Phys.
  Rev. \textbf{181} (1969), 1251.

\bibitem{m20}
P.O. Lovdin, \emph{Theoretical investigation into some properties of ionic
  crystals}, Ph.D. thesis, Uppsala, 1948.

\bibitem{m24}
G.J. McConville, J.Chem. Phys. \textbf{60} (1974), 4093.

\bibitem{m3}
L.S. Metlov, FTVD (Ukraine) \textbf{11} (2001), no.~3, 121.

\bibitem{m23}
J.F. Ogilvie and F.J. Wang, J. Mol. Struct. \textbf{273} (1992), 277.

\bibitem{m10}
A.~Pollian, J.M. Besson, M.~Grimsditch, and W.A. Grosshans, Phys. Rev. B.
  \textbf{39} (1989), 1332.

\bibitem{m19}
K.~Rosciszewski, B.~Pauls, P.~Fulde, and H.~Stoll, Phys. Rev. B. \textbf{60}
  (1999), 7905.

\bibitem{m5}
T.~Rossler and J.B. Page, Phys. Rev. Lett. \textbf{78} (1997), 1287.

\bibitem{m4}
K.W. Sandusky and J.B. Page, Phys. Rev. B \textbf{50} (1994), 866.

\bibitem{m11}
H.~Shimizu, N.~Saitoh, and S.~Sasaki, Phys. Rev. B \textbf{57} (1998), 230.

\bibitem{m25}
S.V. Sinogeikin and D.J. Bass, Phys. Rev. B. \textbf{59} (1999), 14141.

\bibitem{m16}
J.Jr. Skalyo, V.G. Minkiewicz, and G.~Shirane, Phys. Rev. B \textbf{6} (1972),
  4766.

\end{thebibliography}
\end{document}